\documentclass{article} 
\usepackage[final]{colm2026_conference}

\usepackage{amsmath}
\usepackage{microtype}
\usepackage{hyperref}
\usepackage{url}
\usepackage{booktabs}
\usepackage{longtable}
\usepackage{multirow}
\usepackage{tikz}
\usepackage{pgfplots}
\pgfplotsset{compat=1.18}

\usepackage{lineno}

\definecolor{darkblue}{rgb}{0, 0, 0.5}
\hypersetup{colorlinks=true, citecolor=darkblue, linkcolor=darkblue, urlcolor=darkblue}

\title{TAGGRAPH: Tag-Augmented Graphs for Graph Retrieval of Agent Persistent Histories}

\author{Yu-Shu Chen\thanks{ Equal contribution.} \quad Yu-Jung Liang\footnotemark[1] \quad Pengtao Xie \\
Department of Electrical and Computer Engineering\\
University of California, San Diego\\
La Jolla, CA 92093, USA \\
\texttt{\{yuc067,amliang,p1xie\}@ucsd.edu}
}

\begin{document}

\ifcolmsubmission
\linenumbers
\fi

\maketitle
\lhead{Presented at the 2nd Workshop on Lifelong Agents: Learning, Aligning, and Evolving (COLM 2026)}

\begin{abstract}
Long-term memory lets LLM agents recall past interactions and remain consistent across sessions, but memory systems are hard to compare because they often vary in representation, indexing, retrieval, and evaluation. We present a controlled evaluation framework based on shared 5W-style conversational memories. Localized graph configurations traverse a common base graph; AdaptiveGraph adds chronological edges and Personalized PageRank diffusion. We also evaluate BM25 over the same extracted notes and OpenClaw as a raw-input external reference. Retrieval rankings vary across memory settings. On LongMemEval-S, AdaptiveGraph is the strongest graph configuration at 0.844 MRR, but BM25 reaches 0.867 and OpenClaw 0.880. On ATANT Core, localized graph traversal outperforms diffusion and BM25, whereas BM25 leads the stress rounds. Reducing LongMemEval-S within the tested range does not reproduce the ATANT diffusion penalty, but the smallest tested store remains larger than ATANT Core, so store size cannot be ruled out. The penalty also persists under a permissive content-match criterion. Vocabulary normalization and extraction quality substantially affect graph retrieval, and missing extraction tags are common among top-five misses. Retrieval strategies should therefore be evaluated jointly with the memory setting and against strong lexical baselines.

\end{abstract}

\section{Introduction}

LLM agents need memory beyond a fixed context window to personalize responses, answer questions about past events, and remain consistent across sessions. Larger contexts are expensive and bounded. Retrieval-based memory scales by selecting a small set of past interactions from an external store, but a relevant memory that is poorly represented or ranked remains unavailable to the agent. Retrieval quality therefore depends on representation, indexing, and search together.

Memory systems use dense vectors, hybrid lexical-semantic search, or graphs. Embeddings capture semantic similarity but leave entity relations, concept hierarchies, and temporal links implicit. Graphs expose these connections and support retrieval beyond direct word overlap, but they also introduce choices about extraction, topology, traversal, and score propagation. Published systems often change all of these components together, obscuring which choice produces a gain and whether the graph improves on a strong lexical baseline.

We address this problem with a common \textbf{remember()} and \textbf{query()} interface and a shared 5W-style representation based on Who, What, When, Where, Why, and Extra tags. The graph configurations use the same extracted notes, ontology, and base tag-topic-file graph. Localized variants traverse that graph directly; AdaptiveGraph adds chronological file edges and applies Personalized PageRank \citep{page1999pagerank,haveliwala2002topic}. Intuitively, diffusion starts from query-matched graph nodes and spreads relevance along edges, allowing indirectly connected memories to score highly. Our controlled claims concern configurations over this base representation, not graph memory systems in general. OpenClaw \citep{openclaw2026}, which indexes raw text, remains an external reference rather than a controlled ablation.

We evaluate direct tag matching, vocabulary-normalized and TF-IDF-weighted traversal, AdaptiveGraph, and BM25 on LongMemEval \citep{wu2025longmemevalbenchmarkingchatassistants} and ATANT \citep{tanguturi2026atantevaluationframeworkai}. LongMemEval supplies long, distractor-heavy histories; ATANT uses compact persistent narratives and separate stress rounds. AdaptiveGraph leads the graph configurations on LongMemEval-S, but BM25 over the same notes and raw-input OpenClaw score higher. Localized traversal leads on ATANT Core, while BM25 leads the stress rounds. We call the ATANT diffusion penalty ``topic dilution'' only as a descriptive hypothesis: store-size and evaluation-strictness checks do not identify its cause (Section~\ref{sec:discussion-dilution}). Extraction quality also changes the graph results substantially. Together, these findings show that no method performs best across all evaluated settings.

We contribute: (1) a controlled interface for comparing retrieval configurations over shared conversational memories; (2) an interpretable 5W-style base representation supporting localized and diffusion-based retrieval; and (3) an empirical analysis showing that graph and lexical methods trade places across memory settings, with extraction quality constraining all graph configurations.

All code and dataset generation scripts will be provided in a GitHub repository.

\section{Related work}

Prior work varies in both memory representation and retrieval. Surveys of agent memory and graph techniques map a design space spanning storage, consolidation, indexing, and recall \citep{hu2025memoryageaiagents,bei2025graphsmeetaiagents}. We focus on the narrower question of retrieval behavior under a shared conversational representation.

\textbf{Unstructured memory management.}
MemGPT \citep{packer2024memgptllmsoperatingsystems} manages context as virtual memory, MemoryBank \citep{zhong2023memorybankenhancinglargelanguage} adds consolidation and forgetting, and Generative Agents \citep{park2023generativeagentsinteractivesimulacra} retrieve free-text reflections by recency, importance, and embedding relevance. These systems establish the value of persistent memory and selective recall. These systems primarily organize memory through free text, reflections, or embedding-based retrieval. Our experiments examine a complementary representation in which 5W tags provide explicit graph retrieval keys while the original interaction remains the retrieved unit.

\textbf{Embedding-centric production systems.}
Mem0 \citep{chhikara2025mem0buildingproductionreadyai} distills conversations into entries retrieved by embedding similarity, following retrieval-augmented generation \citep{lewis2021retrievalaugmentedgenerationknowledgeintensivenlp}. Dense retrieval handles paraphrase well and is practical at scale, but its memory organization and selection paths are difficult to inspect or edit. OpenClaw \citep{openclaw2026} provides an external production reference that ranks raw text with embedding similarity and BM25 (Section~\ref{sec:method-openclaw}); BM25 over the extracted notes supplies the controlled lexical comparison.

\textbf{Graph-structured memory.}
GraphRAG \citep{edge2025localglobalgraphrag} combines entity graphs with community summaries, HippoRAG \citep{gutierrez2025hipporag} applies Personalized PageRank over OpenIE triples, Zep \citep{rasmussen2025zeptemporalknowledgegraph} maintains a temporal knowledge graph, and MAGMA \citep{jiang2026magma} traverses semantic, temporal, and causal views. A-MEM \citep{xu2025amemagenticmemoryllm} continually restructures stored experiences. Because these systems jointly vary graph representation and retrieval, we compare localized traversal, normalized and TF-IDF-weighted tag retrieval, and PageRank diffusion over one shared 5W representation.

\textbf{Benchmarks and controlled comparison.}
LongMemEval \citep{wu2025longmemevalbenchmarkingchatassistants} uses typed questions over long multi-session histories; ATANT \citep{tanguturi2026atantevaluationframeworkai} tests continuity in shorter narratives. Their different memory regimes let us test whether one ranking strategy transfers across benchmarks. Our graph configurations share ingestion, representation, and base graph construction. AdaptiveGraph's additional chronological edges are explicit and separately ablated, while OpenClaw remains outside this controlled comparison.

\section{Methodology}

Our framework standardizes ingestion, storage, retrieval, and evaluation so that retrieval configurations can be compared under one protocol.

Each backend implements \textbf{remember()} for storage and \textbf{query()} for retrieval. The graph configurations share extracted notes, a 5W ontology, and a base tag-topic-file graph. Localized configurations traverse this graph directly; AdaptiveGraph adds chronological file edges and Personalized PageRank.


\subsection{Memory ingestion and extraction}
\label{sec:method-ingestion}

Each user-assistant interaction is stored as a single Markdown document containing the user prompt and assistant response. These Markdown files serve as the atomic retrieval units, preserving the original dialogue while allowing multiple retrieval algorithms to operate over the same memory collection.

During ingestion, each conversation is processed by a language model that extracts a structured \textbf{5W+Extra} representation, following the journalistic 5W event frame used previously for automated event extraction \citep{hamborg2019giveme5w1huniversalextractingmain}. The extracted fields are \textbf{Who}, \textbf{What}, \textbf{When}, \textbf{Where}, \textbf{Why}, \textbf{Extra}, and a short \textbf{Summary}; the complete schema and extraction prompt are provided in Appendix~\ref{app:prompt}. The 5W fields capture the participants, actions, time, location, motivation, and additional factual details of an interaction in a compact form suitable for graph construction and retrieval.

Extraction is designed for memory retrieval rather than general summarization. It favors precision over recall, extracts only information clearly supported by the interaction, and avoids inferring user identity, preferences, skill level, or long-term goals. User statements are treated as the primary memory source, while assistant responses are used only to clarify the topic and are not used to introduce new factual memories.

Extracted concepts are represented as short hierarchical ontology tags rather than complete factual statements. For example, \texttt{career/software\_engineering}, \texttt{company/google}, and \texttt{programming/python} are stored as separate ontology paths instead of one long label. The model is prompted to prefer an existing ontology entry and to introduce a new tag only when no suitable entry exists; this preference is not enforced after generation. The summary provides a human-readable description of the interaction, while the ontology tags form the primary semantic representation used for graph construction and retrieval. The atomic tags are used only as retrieval keys. The retrieved unit remains the complete Markdown interaction, so decomposing a fact into tags does not remove the original relations, actions, numerical details, or constraints from the context returned to the agent.

\subsection{Graph construction}
\label{sec:method-graph}

Graph configurations use a hierarchical ontology and an entity-topic knowledge graph.

The ontology maintains a hierarchical taxonomy of semantic concepts extracted from conversations. Tags are represented as slash-separated paths encoding parent-child relationships, such as \texttt{career/software\_engineering}, \texttt{programming/python}, and \texttt{location/san\_diego}. As new conversations are ingested, the model receives the current ontology in its prompt and is asked to reuse existing entries when possible. This prompting encourages semantic consistency across the memory collection but does not enforce reuse or eliminate near-duplicate labels. A real excerpt of the ontology is shown in Appendix~\ref{app:ontology}.

The directed graph has four node types: \textbf{tag} nodes for extracted concepts, \textbf{topic} nodes grouping related memories, \textbf{file} nodes for Markdown interactions, and \textbf{ontology} nodes for the taxonomy. Tag-topic and topic-file edges link concepts to memories; similarity edges link topics sharing at least three attributes across the extracted 5W categories; ontology edges encode parent-child relations. These paths let a query move from an ontology match to its associated interactions or to nearby concepts. AdaptiveGraph additionally links consecutive file nodes chronologically.

\subsection{Retrieval methods}

The framework evaluates the original OpenClaw retrieval pipeline, four graph-based retrieval configurations, and a graph-free BM25 baseline that ranks the same extracted notes without touching the graph (Section~\ref{sec:experiment-methods}). The graph-based methods share the same ingestion pipeline, storage format, 5W ontology, and base tag-topic-file graph. The TagGraph configurations retrieve from that base graph by localized traversal, whereas AdaptiveGraph adds chronological file edges to its transition graph, seeds both tag and file nodes, and ranks with Personalized PageRank. We therefore compare retrieval configurations over a shared base representation rather than claim a pure retrieval-algorithm ablation.

\subsubsection{OpenClaw}
\label{sec:method-openclaw}

OpenClaw indexes raw conversation text using dense cosine similarity and BM25 with weights 0.7 and 0.3. It is an external production reference, not a controlled ablation, because it uses neither the extracted store nor the graph. Appendix~\ref{app:implementation} gives full retrieval details.

\subsubsection{BM25}

The controlled lexical baseline applies Okapi BM25 to the full text of the same extracted notes used by the graph configurations. It uses no ontology or graph edges. Unlike OpenClaw, this baseline holds the memory store fixed, directly testing whether graph structure improves on lexical ranking of identical retrieval units. Tokenization and parameters are given in Appendix~\ref{app:implementation}.

\subsubsection{TagGraph}

TagGraph performs local ontology-guided traversal. Lexically matched tags are expanded by bounded breadth-first search over parent and child concepts. Traversing $\text{Tag} \rightarrow \text{Topic} \rightarrow \text{File}$ paths produces candidate files, which are scored by graph distance so that direct matches outrank more remote ontology neighbors.

\textbf{TagGraph-Basic} uses raw lexical overlap. \textbf{TagGraph-Stopwords} removes query tokens occurring in more than 15\% of a case's files before ontology matching, preventing common words from opening broad graph regions. \textbf{TagGraph-TFIDF} also weights content overlap by corpus-wide TF-IDF so that rarer terms contribute more to ranking. The traversal and base graph remain unchanged across these variants.

\subsubsection{AdaptiveGraph}

AdaptiveGraph augments the base graph with chronological file edges and replaces localized traversal with Personalized PageRank \citep{page1999pagerank,haveliwala2002topic}, also known as Random Walk with Restart. Rather than restricting retrieval to direct ontology paths, AdaptiveGraph builds a transition graph containing ontology relationships, topic-file connections, topic similarity edges, and chronological links between consecutive conversation files.

Given a query, matched ontology tags initialize the teleportation vector, together with TF-IDF content-overlap scores assigned to candidate file nodes. Retrieval is then performed using Personalized PageRank, which iteratively updates node relevance according to
\begin{equation*}
x^{(k+1)}=\alpha M^\top x^{(k)}+(1-\alpha)s,
\end{equation*}
where $M$ is the transition matrix, $s$ is the teleportation vector derived from the query, and $\alpha$ is the damping factor. Conversation files are ranked according to their final PageRank probabilities.

Unlike bounded traversal, this update can transfer relevance through topic similarity, ontology, and temporal connections before returning a file score. Direct file seeding also preserves a lexical route when the query does not produce a useful ontology entry point. The component ablation separates diffusion, chronological edges, and file seeding.


\section{Experimental setup}

\subsection{Datasets}
\label{sec:experiment-datasets}


\textbf{LongMemEval} \citep{wu2025longmemevalbenchmarkingchatassistants} contains multi-session conversational histories paired with typed memory questions. We use the 500-question LongMemEval-S split, lightly cleaned of malformed entries. Each question is paired with roughly 40--50 past sessions, of which only one or a few are relevant. We use the provided question types, including multi-session, temporal-reasoning, knowledge-update, and single-session-assistant, for per-type analysis. We also evaluate the oracle variant, where each question is given only the gold sessions, as a distractor-free upper-bound check. 

\textbf{ATANT} \citep{tanguturi2026atantevaluationframeworkai} evaluates conversational continuity across persistent narratives spanning six life domains: Career, Daily Life, Health, Learning, Life Events, and Relationships. We evaluate both the Core split and the Stress Round splits, which draw on separate story sets designed to stress conversational continuity, with more questions per story (Table~\ref{tab:atant-penalty}). 

Together, the two benchmarks let us compare retrieval behavior under different memory structures: LongMemEval provides long, distractor-heavy histories, while ATANT provides compact but evolving personal narratives. 

\subsection{Compared retrieval methods}
\label{sec:experiment-methods}

We compare six retrieval methods: \textbf{OpenClaw}, \textbf{TagGraph-Basic}, \textbf{TagGraph-Stopwords}, \textbf{TagGraph-TFIDF}, \textbf{AdaptiveGraph}, and \textbf{BM25}. The four graph-based methods share the same memory ingestion pipeline, ontology construction procedure, Markdown storage format, and base 5W graph. The localized configurations traverse the base graph directly, whereas AdaptiveGraph adds chronological file edges and applies Personalized PageRank; its components are isolated in the ablation in Appendix~\ref{app:ablation-qa}. BM25 ranks the same extracted notes by Okapi BM25 over the full note text, with no ontology and no graph, which makes it a graph-free lexical baseline on an identical memory store.

OpenClaw is an external production baseline: it indexes raw input directly, without the shared 5W pipeline, ontology, or knowledge graph, so we treat it as a deployed hybrid dense-plus-BM25 reference point rather than a controlled graph-configuration ablation. Section~\ref{sec:method-openclaw} gives its retrieval mechanism.

\subsection{Memory extraction models}
\label{sec:experiment-extractors}

For graph-based methods, each conversation is ingested into a structured 5W+Extra representation. To evaluate extraction robustness, we experiment with three instruction-tuned models: \textbf{Gemma} \citep{gemmateam2025gemma3technicalreport}, \textbf{GPT-OSS} \citep{openai2025gptoss120bgptoss20bmodel}, and \textbf{Qwen3-Small} \citep{qwen3_2025}.
The full prompt is provided in Appendix~\ref{app:prompt}. Holding the prompt fixed allows us to attribute differences in graph-based retrieval performance primarily to the extraction model and retrieval configuration rather than to prompt variation.


\subsection{Evaluation metrics}
\label{sec:experiment-metrics}

Retrieval performance is evaluated primarily with \textbf{Mean Reciprocal Rank (MRR)}. For LongMemEval, retrieved notes are mapped to source sessions, de-duplicated through depth 40, and matched to gold session labels; we also report per-type recall@$k$. Because ATANT lacks gold session labels, a note is relevant when it contains all expected answer substrings, case-insensitive.

\section{Results}

\begin{table}[t]
\centering
\caption{Retrieval MRR across benchmarks and extraction models. Basic, Stop, and TFIDF are localized TagGraph configurations; Adapt is AdaptiveGraph. BM25 ranks the shared extracted notes without a graph, while OpenClaw indexes raw conversations as an external reference. Bold marks the best method on the shared extracted store. Stress is the mean over R2--R5.}
\label{tab:main-results}
\small
\begin{tabular}{llcccccc}
\toprule
Benchmark & Extraction & Basic & Stop & TFIDF & Adapt & BM25 & OpenClaw (orig.) \\
\midrule
LongMemEval-S & GPT-OSS     & 0.296 & 0.404 & 0.428 & 0.768 & \textbf{0.794} & 0.880 \\
LongMemEval-S & Qwen3-Small & 0.263 & 0.419 & 0.435 & 0.817 & \textbf{0.847} & 0.880 \\
LongMemEval-S & Gemma       & 0.495 & 0.706 & 0.718 & 0.844 & \textbf{0.867} & 0.880 \\
LongMemEval (oracle) & all three   & 1.000 & 0.996--0.998 & 1.000 & 1.000 & 1.000 & 0.990 \\
\midrule
ATANT (Core) & GPT-OSS     & 0.682 & \textbf{0.703} & 0.691 & 0.544 & 0.641 & 0.705 \\
ATANT (Core) & Qwen3-Small & 0.729 & \textbf{0.734} & 0.708 & 0.554 & 0.631 & 0.705 \\
ATANT (Core) & Gemma       & 0.677 & \textbf{0.695} & 0.682 & 0.554 & 0.640 & 0.705 \\
ATANT (Stress R2--R5) & GPT-OSS     & 0.815 & 0.821 & 0.827 & 0.821 & \textbf{0.866} & 0.847 \\
ATANT (Stress R2--R5) & Qwen3-Small & 0.812 & 0.816 & 0.823 & 0.809 & \textbf{0.872} & 0.847 \\
ATANT (Stress R2--R5) & Gemma       & 0.813 & 0.822 & 0.824 & 0.826 & \textbf{0.868} & 0.847 \\
\bottomrule
\end{tabular}
\end{table}

Table~\ref{tab:main-results} compares three regimes. Within a row, the graph configurations and BM25 use the same extracted notes; OpenClaw instead indexes raw conversations and is not part of the controlled store comparison. The graph configurations also share the 5W ontology and base graph, although AdaptiveGraph adds chronological edges and changes seeding and ranking.

On LongMemEval-S, AdaptiveGraph is the strongest graph configuration under every extractor, reaching 0.844 MRR with Gemma. BM25 over those notes reaches 0.867, and raw-input OpenClaw reaches 0.880. Thus, diffusion substantially improves on localized traversal without surpassing the lexical baseline. The oracle rows reach at least 0.996 MRR on the extracted store, showing that the main differences arise when distractors must be ranked rather than from failure to parse the gold sessions.

ATANT changes the ordering. TagGraph-Stopwords leads on Core, while AdaptiveGraph ranks last and BM25 lies between diffusion and localized traversal. BM25 then leads the R2--R5 stress average under all extractors. AdaptiveGraph nevertheless approaches the localized configurations as the rounds progress, reaching or exceeding them in two of three R5 settings (Appendix~\ref{app:atant-stress}).

\subsection{Effect of graph diffusion}
\label{sec:results-diffusion}

AdaptiveGraph improves over TagGraph-TFIDF by 0.126--0.382 MRR on LongMemEval-S, yet BM25 remains 0.023--0.030 higher. AdaptiveGraph also outperforms the localized configurations on all LongMemEval question types under GPT-OSS and Qwen3-Small extraction and on five of six types under Gemma; single-session preference is the exception (Appendix~\ref{app:per-type}). Diffusion therefore provides a broad gain over localized traversal, but much of the full configuration's strength is lexical.

The component ablation disables diffusion, chronological edges, or direct TF-IDF file seeding while holding the remaining configuration fixed. Under GPT-OSS extraction, removing file seeding costs 0.119 MRR on LongMemEval-S, compared with 0.020 for diffusion and 0.001 for chronological edges. These leave-one-out effects are marginal and non-additive: diffusion becomes much more valuable when file seeding is absent (Section~\ref{sec:discussion-dilution}). On ATANT Core, the sign reverses, and removing diffusion raises MRR by 0.158. Appendix~\ref{app:ablation-qa} reports complete component values and scoring-implementation checks.

\subsection{Effect of vocabulary normalization}
\label{sec:results-vocab}

Localized traversal is highly sensitive to its lexical graph entry points. Stopword filtering improves LongMemEval-S MRR by 36.5--59.3\% across extractors. Tokens such as ``how'', ``use'', and ``get'' otherwise match broad ontology regions and introduce unrelated files. Filtering improves recall@5 for all six question types under every extractor (Appendix~\ref{app:per-type}).

TF-IDF weighting adds a smaller but consistent gain after filtering, raising MRR from 0.404--0.706 to 0.428--0.718 across the three extractors. Thus, index normalization can matter as much as traversal choice: structural search cannot compensate for noisy query-to-ontology matches.

\subsection{Performance across extraction models}
\label{sec:results-extraction}

Extraction changes performance more than the full localized-retrieval progression. With TagGraph-TFIDF on LongMemEval-S, Gemma reaches 0.718 MRR versus 0.428 for GPT-OSS, a 0.290 difference; the entire Basic-to-TFIDF progression under GPT-OSS adds 0.132. Retrieval comparisons within one extracted store therefore do not remove extraction as a system-level constraint.

The tag vocabularies offer one explanation. Gemma produces 376.5 distinct tags per case, compared with 918.5 for GPT-OSS and 1,140.5 for Qwen3-Small, and reuses each tag more often (Appendix~\ref{app:tag-vocab}). A smaller, more convergent vocabulary may create fewer noisy entry points, but we do not vary convergence independently and therefore do not claim it as the cause.

The extractors also have different per-type blind spots (Appendix~\ref{app:per-type}). GPT-OSS is weakest on assistant-side questions, whereas Qwen3-Small is weakest on knowledge updates. Because the shared prompt asks models not to extract facts found only in assistant responses, assistant-side performance also reflects compliance with a deliberately restrictive ingestion rule (Appendix~\ref{app:prompt}).

\subsection{From ranking to answer accuracy}
\label{sec:results-qa}

We additionally pass each method's top-five notes to the same Qwen3-Small generator on 487 LongMemEval-S cases. Containment accuracy---whether the normalized gold answer appears in the generated answer---follows the retrieval ordering, increasing from 14.6\% for TagGraph-Stopwords to 18.7\% for TagGraph-TFIDF and 25.3\% for AdaptiveGraph. Token F1 gives the same ordering. Because this directional check uses a slightly different retrieval-scoring implementation, full results and metric details appear in Appendix~\ref{app:ablation-qa}.

\subsection{Failure case diagnostics}
\label{sec:results-failure}

We audit whether top-five misses coincide with missing tag-based entry points. A helpful tag is attached to the gold note and matched to the query independently of the retrieval outcome. The pool contains every case missed by at least one graph configuration under at least one extractor, giving 658 gold notes across LongMemEval-S and ATANT.

Of these, 651 lack a helpful tag under at least one extractor; only seven have helpful tags under all extractors but configuration-dependent outcomes. The unit is the gold note, which is stricter than LongMemEval's session-level metric: another note from the same gold session may rank highly while this diagnostic records a miss. The counts are therefore not directly comparable with the main recall values. This pooled analysis audits reachability rather than causally decomposing all retrieval errors; category definitions and examples appear in Appendix~\ref{app:failure-examples}.

\section{Discussion}

\subsection{When diffusion helps and when it hurts}
\label{sec:discussion-dilution}

No graph configuration wins across settings. AdaptiveGraph leads on LongMemEval-S, where diffusion can propagate relevance to indirectly connected memories; localized traversal leads on ATANT Core.

A controlled LongMemEval-S sweep keeps the gold sessions and samples 0, 3, 8, 18, or 36 distractor sessions. Diffusion contributes 0.000 MRR without distractors and about $+0.02$ at higher densities; the contribution is non-negative but not monotonic across the tested range. All three extractors reproduce this pattern. With only gold sessions, every candidate is correct, and ordering cannot affect MRR. The sweep supports a distractor-dependent benefit within LongMemEval-S, not a general claim that more distractors monotonically increase diffusion's value.

Diffusion overlaps with direct file seeding. With seeding disabled at 36 distractors, adding diffusion raises MRR from 0.101 to 0.678, a $+0.577$ contribution compared with $+0.020$ when seeding is present. The two are strongly substitutable in aggregate MRR under this GPT-OSS configuration; we did not test whether they recover the same cases. BM25’s LongMemEval lead reinforces the importance of this lexical route: diffusion improves localized traversal, while direct file seeding is the largest marginal contributor when both components are present.

On ATANT Core, removing diffusion raises MRR by 0.158. We call this penalty ``topic dilution'' only as a descriptive hypothesis: propagation may spread probability across similarly connected memories in a compact graph, but we measure neither topic concentration nor the proposed score redistribution. The demonstrated result is the ranking penalty, not its mechanism.

Reducing LongMemEval-S to 24.3 notes per case does not reproduce the penalty, but the sweep does not reach ATANT Core's 7.7 notes per story. Store size therefore does not explain the effect within the tested LongMemEval-S range but cannot be ruled out. Across separate ATANT story sets, the penalty generally shrinks with store size, though not monotonically (Table~\ref{tab:atant-penalty}); this comparison does not isolate size.

Evaluation strictness also does not account for the penalty: under permissive any-substring matching, removing diffusion still improves MRR by 0.135--0.143 across extractors (Appendix~\ref{app:strict-any}).

The reproducible result is the ranking penalty; its cause remains unidentified, and the ATANT stress-round crossover shows that it is not fixed to the benchmark (Appendix~\ref{app:atant-stress}). These results motivate comparing graph retrieval with lexical ranking in the target setting before choosing how aggressively to diffuse.

\subsection{Extraction constrains graph retrieval}

Missing lexical tag entry points are common among top-five misses. For localized configurations, an absent helpful tag removes the direct graph path to a note. AdaptiveGraph can still seed file nodes from TF-IDF overlap, so missing tags restrict but do not preclude recovery. Because the diagnostic is conditioned on a pooled miss set and has no successful-case control group, it cannot estimate the causal contribution of extraction failure.

The extraction prompt prioritizes user statements and uses assistant responses mainly for context. This avoids inventing user facts but can drop assistant-side evidence. For example, the extractors omit an assistant recommendation needed by a LongMemEval question about a restaurant in Cihampelas.

Potential improvements include validation before graph insertion, selective preservation of assistant-provided facts, and more consistent handling of synonyms and subtopics.

\subsection{Limitations and future work}

LLM extraction variance propagates through every graph result. The implementation rebuilds each case graph rather than updating it incrementally, and flat Markdown storage is not designed for deployment scale. All graph configurations inherit the 5W schema; because we do not compare free-text summaries or relation triples, their ranking may be representation-specific. The ATANT crossover also remains unexplained: strictness does not account for it, and store size is only partially tested. Identifying the cause requires direct measurement of graph structure and score redistribution.

Our primary metric is retrieval ranking; Section~\ref{sec:results-qa} adds only a directional answer-accuracy check. Dense and hybrid retrievers beyond the external OpenClaw reference, stronger graph-memory systems such as HippoRAG, Zep, MAGMA, and A-MEM, frontier-model extraction, cost, and full LLM-judge evaluation remain outside scope. AdaptiveGraph's power-iteration scores are also less traceable than TagGraph's ontology paths.

\bibliography{content/references}
\bibliographystyle{colm2026_conference}

\appendix
\section{Example 5W Memory Note}
\label{app:5w-example}

Each conversation turn is stored as one markdown note: a tag line holding the
extracted 5W ontology paths, followed by the raw user/assistant exchange. The
tag line is the semantic index the graph is built from
(Section~3.2); the body is preserved verbatim for content scoring and
auditability. The example below is taken unmodified from the LongMemEval-S
corpus (Qwen3-Small extraction; assistant response truncated for space).

\begin{verbatim}
#who:user #what:activity/yoga #what:health/sleep
#what:topic/yoga_poses #when:time/morning
#why:goal/sleep_quality #where:n/a

### User
I've been doing some yoga in the morning, and I'm curious
to know if there are any specific yoga poses that can help
improve my sleep quality.

---

### Assistant
Yoga is an excellent way to start the day! Yes, certain
yoga poses can indeed help improve sleep quality. Here are
some sleep-promoting yoga poses you can incorporate into
your morning routine:

1. **Child's Pose (Balasana)**: This gentle, relaxing pose
   can help calm your mind and body [...]
\end{verbatim}

Every tag is a slash-delimited ontology path in one of the 5W categories
(\texttt{who}, \texttt{what}, \texttt{when}, \texttt{where}, \texttt{why},
plus \texttt{extra}); \texttt{n/a} marks categories with no explicitly
supported value, and no fact is inferred beyond what the turn states. Tag
nodes link this note into the tag$\to$topic$\to$file graph traversed at
retrieval time.


\section{Extraction Prompt}
\label{app:prompt}

The complete 5W+Extra extraction prompt is reproduced below. All three
extraction models (Gemma, GPT-OSS, Qwen3-Small) receive this same
prompt at temperature 0; the placeholders \texttt{\{tags\}},
\texttt{\{text\_user\}}, and \texttt{\{text\_assistant\}} are filled per
conversation turn with the current ontology tag list and the turn's raw
text. Using one shared prompt is what lets Section~\ref{sec:results-extraction}
trace performance differences to the extraction models themselves rather than
to prompt engineering.

\begin{small}
\begin{verbatim}
Extract memory units from the following interaction into a
5W+Extra structure.

Return strictly valid JSON with the following schema:

{
  "who": [], "what": [], "when": [], "where": [],
  "why": [], "extra": [], "summary": ""
}

### Purpose

The goal is to generate retrieval-oriented tags for a memory
graph. Tags should help future memory retrieval, clustering,
and relationship discovery. Tags are not natural language
descriptions.

### Extraction Principles

1. Extract only information explicitly supported by the
   interaction.
2. Do not infer user preferences, beliefs, identity, expertise,
   goals, employment, education, or long-term interests unless
   explicitly stated.
3. A user asking about a topic does not imply interest,
   ownership, expertise, intent, or preference.
4. Explicit user facts may be extracted even when they are not
   the primary topic of the interaction, provided they could be
   useful for future retrieval.
5. User statements are the primary source for memory extraction.
6. Assistant responses may be used only to clarify the topic
   being discussed.
7. Do not extract facts that appear only in the assistant
   response.
8. Extract the topics discussed in the interaction, even when
   they are not user facts.
9. Prioritize precision over recall. If uncertain, omit the tag.

### Tag Rules

1. Use short, atomic tags.
2. Prefer multiple simple tags over one highly specific tag.
3. Use hierarchical paths whenever possible.
4. Prefer existing ontology tags before creating new ones.
5. Use lowercase only.
6. Use underscores instead of spaces.
7. Prefer singular nouns.
8. Keep hierarchy depth between 1 and 4 levels.
9. Do not encode complete facts into tag names.

Good:
    career/software_engineering
    specialization/backend
    programming/python
    company/google

Bad:
    user_is_a_backend_python_engineer_at_google

### 5W Definitions

#### who
People, organizations, teams, identities, or actors explicitly
mentioned.
Examples: user, person/john_smith, company/google,
organization/openai

#### what
Topics, activities, projects, products, domains, concepts,
technologies, events, or subjects discussed.
Examples: programming/python, project/openclaw,
career/software_engineering, hobby/photography
If multiple distinct topics are discussed, include multiple
tags.

#### when
Explicit dates, time periods, ages, timelines, schedules, or
temporal references.
Examples: year/2026, month/may, time/childhood

#### where
Explicit locations, regions, countries, cities, workplaces, or
geographic references.
Examples: country/japan, city/san_diego, location/home

#### why
Explicit motivations, reasons, goals, intentions, or causes
stated in the interaction.
Examples: goal/career_growth, reason/cost_saving,
motivation/learning
Only include motivations that are directly stated.

#### extra
Specific factual details, attributes, qualifications,
possessions, preferences, characteristics, or background
information explicitly stated by the user that do not naturally
belong in the other 5W categories.
Examples: degree/business_administration, language/japanese,
pet/cat, device/keychron_q1

### Summary

Write a single sentence describing the interaction.
Requirements:
* Describe what was discussed.
* Do not infer user traits.
* Do not invent facts.
* Keep under 30 words.

Good:
    The interaction discussed Rust programming and its learning
    curve.
Bad:
    The user is interested in becoming a Rust developer.

### Empty Categories

If a category has no valid tags, return an empty array:

{
  "who": [], "what": [], "when": [], "where": [],
  "why": [], "extra": []
}

Do not use null.
Do not use "n/a".

### Existing Ontology Tags

{tags}

### Interaction

User:
{text_user}

Assistant:
{text_assistant}
\end{verbatim}
\end{small}

\section{Ontology Excerpt}
\label{app:ontology}

Each case keeps a growing ontology, saved as an indented markdown
hierarchy (\texttt{ontology.md}) that the extraction prompt receives as its
existing-tag list and that retrieval walks through for tag expansion. The
excerpt below is taken unmodified from a LongMemEval-S case (Qwen3-Small
extraction); the full file for this case spans 1,479 lines and roughly 90
top-level categories. Indentation encodes parent-child edges: for example,
the path \texttt{device/fitbit\_inspire\_hr} appears as the child
\texttt{fitbit\_inspire\_hr} under the top-level category \texttt{device}.

\begin{small}
\begin{verbatim}
- activity
    - birding
    - fitness_tracking
    - photography
    - running
    - swimming
    - tennis
    - volunteering
- career
    - job_search
    - senior_motion_designer
- city
    - san_diego
- device
    - fitbit_inspire_hr
    - fitness_tracker
    - gps_running_watch
    - phone
    - wireless_blood_pressure_monitor
- duration
    - 15_to_30_minutes
    - 30_minutes
    - 45_minutes
- goal
    - fitness
    - sleep_quality
    - step_count
- health
    - blood_pressure
    - sleep
[...]
\end{verbatim}
\end{small}

The prompt asks the extraction model to create a new tag only when no existing ontology entry fits (Section~\ref{sec:method-graph}); the implementation does not enforce this preference after generation. The same requested normalization rules (lowercase, underscores, singular nouns, depth $\leq 4$) are used for every extraction model.

\section{Tag Vocabulary Convergence by Extraction Model}
\label{app:tag-vocab}

Section~\ref{sec:results-extraction} attributes part of Gemma's advantage to a
more convergent tag vocabulary. Table~\ref{tab:tag-vocab} measures that
directly on the LongMemEval-S corpus: for every case we collect the ontology
paths on each note's tag line, drop \texttt{n/a}, and count how many distinct
tags the case uses and how often each one is reused. Cases with no notes are
excluded, which is why the case counts fall below 500.

\begin{table}[h]
\centering
\caption{Tag vocabulary statistics on LongMemEval-S. Reuse is tag occurrences
divided by distinct tags within a case, averaged over cases: a higher value
means the same tags are used again instead of new ones being minted.}
\label{tab:tag-vocab}
\small
\begin{tabular}{lrrrrr}
\toprule
Extraction & Cases & Notes/case & Distinct tags/case & Tags/note & Reuse \\
\midrule
Gemma       & 496 & 249.5 & \textbf{376.5}  & 4.95 & \textbf{3.27} \\
GPT-OSS     & 469 & 232.1 & 918.5           & 6.50 & 1.64 \\
Qwen3-Small & 482 & 240.1 & 1{,}140.5       & 9.29 & 1.96 \\
\bottomrule
\end{tabular}
\end{table}

Gemma names 376.5 distinct concepts per case against 918.5 for GPT-OSS and
1,140.5 for Qwen3-Small, and reuses each one about twice as often, on a
comparable number of notes. This is one observation on one corpus, so it
supports the reading in Section~\ref{sec:results-extraction} without isolating
vocabulary convergence as the cause of the retrieval difference.

\section{Implementation Details}
\label{app:implementation}

The framework is implemented in Python using a shared backend interface
for all retrieval methods. Individual conversation turns are stored as
markdown files containing one user message and one assistant response.
The ontology is stored separately as a hierarchical markdown document, and
the knowledge graph is implemented using NetworkX \citep{hagberg2008networkx}.

OpenClaw embeds queries and raw stored chunks with
nomic-embed-text-v1.5 \citep{nussbaum2025nomicembedtrainingreproducible} and
ranks them by cosine similarity. Its lexical retriever uses Okapi BM25 over
SQLite FTS5: query tokens are quoted and joined conjunctively, and FTS5 rank
is mapped to a bounded relevance score. Each retriever returns four times the
requested depth; candidates are merged by memory identifier and scored as
$0.7\,s_{\mathrm{dense}}+0.3\,s_{\mathrm{BM25}}$, with zero contribution
from a list in which the memory is absent. Maximal marginal relevance and
temporal decay are available but disabled in our benchmark configuration.

All graph-based retrieval methods use the same memory storage, ontology
construction, graph construction, and evaluation code. For TagGraph
retrieval, ontology expansion is performed using bounded breadth-first
traversal. The generic-token threshold for TagGraph-Stopwords and
TagGraph-TFIDF is set to 15\% document frequency within each case. The BM25
baseline uses the Okapi BM25 implementation of \texttt{rank\_bm25} with its
default parameters ($k_1=1.5$, $b=0.75$) over full note text, sharing its
tokenizer with the TF-IDF scorer. For
AdaptiveGraph, Personalized PageRank uses a damping factor of
$\alpha=0.85$ and runs for 20 iterations.

Retrieval depth is fixed at 40 throughout evaluation. We checked the
sensitivity of this choice on one configuration: TagGraph-TFIDF over
LongMemEval-S changes by less than 0.004 MRR across retrieval depths 20, 40,
and 80, and its recall@5 does not change. We did not repeat the sweep for
the other methods or for ATANT.

\section{Component Ablation and Downstream Answer Accuracy}
\label{app:ablation-qa}

Table~\ref{tab:component-ablation} gives the full component ablation from
Section~\ref{sec:results-diffusion}. Each row removes one AdaptiveGraph
component while the graph and the initial query scores stay fixed; setting
the iteration count to zero removes diffusion alone.

\begin{table}[h]
\centering
\caption{Change in MRR from removing one AdaptiveGraph component
(GPT-OSS extraction). Negative means the component was contributing;
positive means removing it improves accuracy.}
\label{tab:component-ablation}
\small
\begin{tabular}{lccc}
\toprule
Dataset & $-$Diffusion & $-$Chronological & $-$File seeding \\
\midrule
LongMemEval-S & $-0.020$ & $-0.001$ & $\mathbf{-0.119}$ \\
ATANT (Core)  & $\mathbf{+0.158}$ & $+0.077$ & $-0.009$ \\
\bottomrule
\end{tabular}
\end{table}

The ablation uses the scoring implementation from the downstream
answer-accuracy run, which differs slightly from the implementation behind
Table~\ref{tab:main-results}; shared ATANT Core configurations agree within
0.007 MRR. As a cross-implementation check, AdaptiveGraph without diffusion
scores 0.705 on ATANT Core, while the independently implemented
TagGraph-Stopwords scores 0.703. They remain algorithmically distinct:
TagGraph ranks by graph distance, whereas AdaptiveGraph without propagation
ranks by its teleportation vector.

\subsection{Strict and Permissive Content Match on ATANT}
\label{app:strict-any}

ATANT provides no gold session labels, so relevance is judged by content
match. Section~\ref{sec:discussion-dilution} asks whether the strict
criterion, which requires every expected substring, is what makes diffusion
harmful on ATANT Core. Table~\ref{tab:strict-any} recomputes the diffusion
penalty under the permissive criterion, which requires any expected
substring, from the same per-instance runs. The penalty shrinks by about a
tenth and stays above $+0.13$ under all three extraction models, so the
strictness of the criterion does not account for it.

\begin{table}[h]
\centering
\caption{ATANT Core MRR with and without diffusion under both content-match
criteria ($n=304$ questions). Penalty is the MRR gain from removing
diffusion, computed before rounding; MRR values use the scoring
implementation of Section~\ref{sec:results-diffusion} and agree with
Table~\ref{tab:main-results} within 0.007.}
\label{tab:strict-any}
\small
\begin{tabular}{lccc|ccc}
\toprule
 & \multicolumn{3}{c|}{Strict} & \multicolumn{3}{c}{Permissive} \\
Extraction & Full & $-$Diffusion & Penalty & Full & $-$Diffusion & Penalty \\
\midrule
GPT-OSS     & 0.546 & 0.705 & $+0.158$ & 0.570 & 0.713 & $+0.143$ \\
Qwen3-Small & 0.547 & 0.699 & $+0.153$ & 0.574 & 0.714 & $+0.140$ \\
Gemma       & 0.549 & 0.699 & $+0.150$ & 0.577 & 0.712 & $+0.135$ \\
\bottomrule
\end{tabular}
\end{table}

\subsection{Downstream Answer Accuracy}

Table~\ref{tab:qa-accuracy} gives the full downstream answer-accuracy
results from Section~\ref{sec:results-qa}. The prompt instructs the
generator to answer only from the provided context; the answer-generation
pipeline's validation excludes 13 cases whose haystack lists a session
twice, leaving $N=487$. Token F1 understates correctness because generated
answers are verbose: an answer containing the gold string ``three'' inside
a full sentence scores 0.17. Containment, whether the normalized gold
answer appears inside the prediction, is immune to verbosity but misses
paraphrases. The two err in opposite directions. Both are computed identically for
all three methods and both give the same ordering.
This run uses a second retrieval-scoring implementation. Its MRRs are 0.409
for TagGraph-Stopwords, 0.463 for TagGraph-TFIDF, and 0.773 for AdaptiveGraph,
compared with 0.404, 0.428, and 0.768 in
Table~\ref{tab:main-results}. The TagGraph-TFIDF difference is larger than
the excluded cases explain, so we treat the answer-accuracy result as a
directional check rather than an exact replication of the headline MRR run.

\begin{table}[h]
\centering
\caption{Downstream answer accuracy on LongMemEval-S (GPT-OSS extraction,
Qwen3-Small generation, $N=487$). Containment: the normalized gold answer
appears inside the generated answer.}
\label{tab:qa-accuracy}
\small
\begin{tabular}{lccc}
\toprule
Retrieval & MRR & Token F1 & Containment \\
\midrule
TagGraph-Stopwords & 0.409 & 0.058 & 14.6\% \\
TagGraph-TFIDF     & 0.463 & 0.072 & 18.7\% \\
AdaptiveGraph      & \textbf{0.773} & \textbf{0.086} & \textbf{25.3\%} \\
\bottomrule
\end{tabular}
\end{table}

\section{Per-Question-Type Recall@1/3/5}
\label{app:per-type}

Table~\ref{tab:per-type} gives the full per-question-type recall@1,
recall@3, and recall@5 breakdown on LongMemEval-S behind the headline
numbers in Section~\ref{sec:results-vocab} and
Section~\ref{sec:results-extraction}. The OC-notes column is OpenClaw
retrieving over the same 5W-extracted notes as the four graph
configurations, computed from the per-instance rank data of an OpenClaw run over the
extracted notes. This is a controlled comparison (same memory representation for
every column) and is distinct from the native OpenClaw (orig.) reference in
Table~\ref{tab:main-results}, which reads raw, unextracted conversation text and does not
depend on the extraction model.

\begingroup
\footnotesize
\begin{longtable}{llrccccc}

\label{tab:per-type}\\

\toprule
Question type & Extraction & $k$ & Basic & Stop & TFIDF & Adapt & OC-notes \\
\midrule
\endfirsthead

\multicolumn{8}{c}{{\bfseries Table \thetable\ -- continued from previous page}} \\
\toprule
Question type & Extraction & $k$ & Basic & Stop & TFIDF & Adapt & OC-notes \\
\midrule
\endhead

\midrule
\multicolumn{8}{r}{{Continued on next page...}} \\
\endfoot

\bottomrule
\endlastfoot

\multirow{9}{*}{single-session-user}
 & \multirow{3}{*}{GPT-OSS} & 1 & 24.3 & 30.0 & 34.3 & \textbf{88.6} & 82.9 \\
 &  & 3 & 45.7 & 55.7 & 58.6 & \textbf{98.6} & 97.1 \\
 &  & 5 & 55.7 & 65.7 & 67.1 & \textbf{98.6} & 97.1 \\
\cmidrule(lr){2-8}
 & \multirow{3}{*}{Qwen3-Small} & 1 & 20.0 & 31.4 & 27.1 & \textbf{87.1} & 82.9 \\
 &  & 3 & 37.1 & 60.0 & 58.6 & \textbf{97.1} & 94.3 \\
 &  & 5 & 47.1 & 67.1 & 67.1 & \textbf{97.1} & 94.3 \\
\cmidrule(lr){2-8}
 & \multirow{3}{*}{Gemma} & 1 & 47.1 & 65.7 & 61.4 & \textbf{87.1} & 85.7 \\
 &  & 3 & 68.6 & 90.0 & 92.9 & \textbf{98.6} & 95.7 \\
 &  & 5 & 72.9 & 91.4 & 94.3 & \textbf{98.6} & 95.7 \\
\midrule
\multirow{9}{*}{single-session-preference}
 & \multirow{3}{*}{GPT-OSS} & 1 & 16.7 & 26.7 & 23.3 & \textbf{36.7} & 73.3 \\
 &  & 3 & 33.3 & 36.7 & 40.0 & \textbf{60.0} & 86.7 \\
 &  & 5 & 36.7 & 43.3 & 46.7 & \textbf{66.7} & 93.3 \\
\cmidrule(lr){2-8}
 & \multirow{3}{*}{Qwen3-Small} & 1 & 6.7 & 10.0 & 13.3 & \textbf{36.7} & 73.3 \\
 &  & 3 & 20.0 & 26.7 & 36.7 & \textbf{53.3} & 86.7 \\
 &  & 5 & 23.3 & 33.3 & 43.3 & \textbf{56.7} & 90.0 \\
\cmidrule(lr){2-8}
 & \multirow{3}{*}{Gemma} & 1 & 30.0 & 36.7 & \textbf{46.7} & 40.0 & 66.7 \\
 &  & 3 & 50.0 & 53.3 & \textbf{66.7} & 56.7 & 86.7 \\
 &  & 5 & 53.3 & 60.0 & \textbf{66.7} & 60.0 & 90.0 \\
\midrule
\multirow{9}{*}{single-session-assistant}
 & \multirow{3}{*}{GPT-OSS} & 1 & 12.5 & 30.4 & 28.6 & \textbf{67.9} & 67.9 \\
 &  & 3 & 26.8 & 48.2 & 48.2 & \textbf{67.9} & 67.9 \\
 &  & 5 & 33.9 & 50.0 & 48.2 & \textbf{67.9} & 67.9 \\
\cmidrule(lr){2-8}
 & \multirow{3}{*}{Qwen3-Small} & 1 & 3.6 & 21.4 & 21.4 & \textbf{96.4} & 98.2 \\
 &  & 3 & 17.9 & 44.6 & 46.4 & \textbf{98.2} & 98.2 \\
 &  & 5 & 21.4 & 53.6 & 58.9 & \textbf{98.2} & 98.2 \\
\cmidrule(lr){2-8}
 & \multirow{3}{*}{Gemma} & 1 & 32.1 & 57.1 & 58.9 & \textbf{100.0} & 98.2 \\
 &  & 3 & 48.2 & 83.9 & 85.7 & \textbf{100.0} & 100.0 \\
 &  & 5 & 50.0 & 89.3 & 89.3 & \textbf{100.0} & 100.0 \\
\midrule
\multirow{9}{*}{multi-session}
 & \multirow{3}{*}{GPT-OSS} & 1 & 17.3 & 27.8 & 34.6 & \textbf{84.2} & 85.7 \\
 &  & 3 & 46.6 & 56.4 & 57.9 & \textbf{91.0} & 97.7 \\
 &  & 5 & 55.6 & 68.4 & 69.9 & \textbf{93.2} & 99.2 \\
\cmidrule(lr){2-8}
 & \multirow{3}{*}{Qwen3-Small} & 1 & 21.1 & 33.1 & 34.6 & \textbf{80.5} & 89.5 \\
 &  & 3 & 40.6 & 59.4 & 63.2 & \textbf{91.0} & 97.7 \\
 &  & 5 & 44.4 & 68.4 & 71.4 & \textbf{94.0} & 99.2 \\
\cmidrule(lr){2-8}
 & \multirow{3}{*}{Gemma} & 1 & 39.1 & 63.2 & 66.2 & \textbf{80.5} & 81.2 \\
 &  & 3 & 59.4 & 80.5 & 82.0 & \textbf{88.0} & 94.7 \\
 &  & 5 & 66.2 & 84.2 & 85.0 & \textbf{90.2} & 95.5 \\
\midrule
\multirow{9}{*}{knowledge-update}
 & \multirow{3}{*}{GPT-OSS} & 1 & 24.4 & 35.9 & 41.0 & \textbf{92.3} & 92.3 \\
 &  & 3 & 51.3 & 61.5 & 66.7 & \textbf{93.6} & 96.2 \\
 &  & 5 & 57.7 & 69.2 & 73.1 & \textbf{94.9} & 96.2 \\
\cmidrule(lr){2-8}
 & \multirow{3}{*}{Qwen3-Small} & 1 & 20.5 & 33.3 & 37.2 & \textbf{65.4} & 67.9 \\
 &  & 3 & 43.6 & 52.6 & 52.6 & \textbf{69.2} & 71.8 \\
 &  & 5 & 47.4 & 56.4 & 56.4 & \textbf{70.5} & 71.8 \\
\cmidrule(lr){2-8}
 & \multirow{3}{*}{Gemma} & 1 & 51.3 & 75.6 & 75.6 & \textbf{91.0} & 96.2 \\
 &  & 3 & 79.5 & 96.2 & 96.2 & \textbf{97.4} & 100.0 \\
 &  & 5 & 80.8 & 96.2 & 96.2 & \textbf{98.7} & 100.0 \\
\midrule
\multirow{9}{*}{temporal-reasoning}
 & \multirow{3}{*}{GPT-OSS} & 1 & 16.5 & 23.3 & 23.3 & \textbf{51.1} & 63.9 \\
 &  & 3 & 33.8 & 42.9 & 43.6 & \textbf{63.9} & 68.4 \\
 &  & 5 & 39.9 & 51.9 & 52.6 & \textbf{66.2} & 69.9 \\
\cmidrule(lr){2-8}
 & \multirow{3}{*}{Qwen3-Small} & 1 & 15.0 & 31.6 & 33.8 & \textbf{75.9} & 87.2 \\
 &  & 3 & 36.1 & 55.6 & 54.9 & \textbf{89.5} & 92.5 \\
 &  & 5 & 45.9 & 60.2 & 63.2 & \textbf{91.0} & 96.2 \\
\cmidrule(lr){2-8}
 & \multirow{3}{*}{Gemma} & 1 & 29.3 & 51.9 & 52.6 & \textbf{70.7} & 82.7 \\
 &  & 3 & 57.1 & 75.9 & 75.2 & \textbf{81.2} & 88.7 \\
 &  & 5 & 61.7 & 78.2 & 76.7 & \textbf{82.7} & 89.5 \\
\end{longtable}
\endgroup

\section{ATANT Core Results by Category}
\label{app:atant-category}

Table~\ref{tab:atant-category} breaks the ATANT Core results (GPT-OSS
extraction) down by the benchmark's six life domains. The diffusion penalty
discussed in Section~\ref{sec:discussion-dilution} appears in every domain rather than being driven by a single domain:
AdaptiveGraph trails every localized variant in all six categories, with the
largest gaps on Career ($0.608$ vs $0.844$ for Stopwords) and Learning
($0.540$ vs $0.753$ for Basic). The localized variants' internal order
varies by category (Stopwords leads on Career and Health, TF-IDF on Daily
Life), matching the small overall spreads in Table~\ref{tab:main-results}.

\begin{table}[h]
\centering
\caption{ATANT Core MRR by category (GPT-OSS extraction). B/S/T =
TagGraph-Basic/-Stopwords/-TFIDF, A = AdaptiveGraph.}
\label{tab:atant-category}
\begin{tabular}{lrcccc}
\toprule
Category & $n$ & B & S & T & A \\
\midrule
Career        &  32 & 0.779 & \textbf{0.844} & 0.823 & 0.608 \\
Daily Life    &  41 & 0.739 & 0.773 & \textbf{0.827} & 0.632 \\
Health        &  31 & 0.685 & \textbf{0.702} & 0.677 & 0.640 \\
Learning      &  36 & \textbf{0.753} & 0.749 & 0.731 & 0.540 \\
Life Events   & 107 & 0.639 & \textbf{0.649} & 0.646 & 0.489 \\
Relationships &  57 & 0.620 & \textbf{0.644} & 0.583 & 0.499 \\
\bottomrule
\end{tabular}
\end{table}

\section{ATANT Stress-Round MRR}
\label{app:atant-stress}

Table~\ref{tab:atant-stress} gives the exact per-split MRR behind the diffusion
penalty crossover discussed in Section~\ref{sec:discussion-dilution}. ``Local'' is the
best of the three localized TagGraph variants (TagGraph-Basic,
TagGraph-Stopwords, TagGraph-TFIDF) for that split and extraction model, and
``Adapt'' is AdaptiveGraph. Bold marks the splits where AdaptiveGraph reaches
or exceeds the best localized variant. The gap on Core is $-0.15$ to $-0.18$
MRR and closes to zero or positive by R5.

\begin{table}[h]
\centering
\caption{ATANT MRR by split, best localized variant versus AdaptiveGraph.}
\label{tab:atant-stress}
\begin{tabular}{lcccccc}
\toprule
 & \multicolumn{2}{c}{GPT-OSS} & \multicolumn{2}{c}{Qwen3-Small} & \multicolumn{2}{c}{Gemma} \\
\cmidrule(lr){2-3}\cmidrule(lr){4-5}\cmidrule(lr){6-7}
Split & Local & Adapt & Local & Adapt & Local & Adapt \\
\midrule
Core & 0.703 & 0.544 & 0.734 & 0.554 & 0.695 & 0.554 \\
R2   & 0.764 & 0.757 & 0.764 & 0.752 & 0.779 & 0.770 \\
R3   & 0.825 & 0.805 & 0.810 & 0.781 & 0.811 & 0.807 \\
R4   & 0.861 & 0.851 & 0.848 & 0.839 & 0.841 & \textbf{0.853} \\
R5   & 0.857 & \textbf{0.872} & 0.869 & 0.863 & 0.864 & \textbf{0.872} \\
\bottomrule
\end{tabular}
\end{table}

Table~\ref{tab:atant-penalty} adds the store statistics and the per-split
diffusion penalty behind Section~\ref{sec:discussion-dilution}. Notes per
story is identical across the three extraction models because each
conversation turn yields one note; the penalty is computed from the same
runs as Table~\ref{tab:strict-any}.

\begin{table}[h]
\centering
\caption{ATANT questions, store size, and diffusion penalty by split.
Penalty is the strict-MRR gain from removing diffusion (iteration count
zero), computed before rounding.}
\label{tab:atant-penalty}
\small
\begin{tabular}{lrrccc}
\toprule
 & & & \multicolumn{3}{c}{Diffusion penalty} \\
\cmidrule(lr){4-6}
Split & Questions & Notes/story & GPT-OSS & Qwen3-Small & Gemma \\
\midrule
Core & 304 & 7.7 & $+0.158$ & $+0.153$ & $+0.150$ \\
R2   & 367 & 6.7 & $+0.075$ & $+0.080$ & $+0.073$ \\
R3   & 386 & 6.2 & $+0.085$ & $+0.094$ & $+0.079$ \\
R4   & 380 & 5.1 & $+0.049$ & $+0.049$ & $+0.049$ \\
R5   & 398 & 4.9 & $+0.035$ & $+0.026$ & $+0.029$ \\
\bottomrule
\end{tabular}
\end{table}

\section{Failure Category Examples}
\label{app:failure-examples}

The diagnostic groups cases as follows: C1, no extraction model supplies a
helpful tag; C2, some models do and some do not; C3, every model supplies a
helpful tag but every retrieval configuration misses; and C4, every model
supplies a helpful tag but only some configurations recover the note. A
helpful tag is attached to the gold note and matched to the query independently
of retrieval outcome.

\begin{table}[h]
\centering
\caption{Failure diagnostics over the 658-case pooled failure subset. Cases are included if at least one of the four graph retrieval configurations under at least one extraction model misses the gold note at recall@5. C1--C4 are defined in the text; examples appear in Appendix~\ref{app:failure-examples}. Model cells give Tagged / Succeeded percentages within the category.}
\label{tab:failure-diagnostics}
\small
\begin{tabular}{lrrccc}
\toprule
Category & $n_{\text{LongMem}}$ & $n_{\text{ATANT}}$ & Gemma & GPT-OSS & Qwen3-Small \\
\midrule
C1 & 398 & 212 & 0.0 / 0.0 & 0.0 / 0.0 & 0.0 / 0.0 \\
C2 &   0 &  41 & 43.9 / 100.0 & 36.6 / 100.0 & 53.7 / 100.0 \\
C3 &   0 &   0 & --- & --- & --- \\
C4 &   0 &   7 & 100.0 / 85.7 & 100.0 / 100.0 & 100.0 / 100.0 \\
\bottomrule
\end{tabular}
\end{table}

One example per category from the diagnostics matrix in
Section~\ref{sec:results-failure} (Table~\ref{tab:failure-diagnostics}).

\textbf{C1, tags missing under all extractors} (case \texttt{e47becba},
LongMemEval-S): question ``What degree did I graduate with?''; none of the
extraction models attaches any tags to the gold note, so no configuration has
a tag-based entry point to it under any extraction model.

\textbf{C2, tag discrepancy} (case \texttt{1\_q1}, ATANT Core):
question ``What company did I interview at?''; GPT-OSS tags the gold note
with \texttt{career/interview}, while Gemma and Qwen3-Small attach no
helpful tag to the same note, so only the GPT-OSS-extracted graph contains
the helpful tag identified by this diagnostic.

\textbf{C4, partial retrieval success} (case \texttt{12\_q2}, ATANT Core):
question ``How many raised beds did I build and what size are they?''; under
Gemma extraction, TagGraph-Basic, TagGraph-Stopwords, and TagGraph-TFIDF all
reach recall@5 $=1.0$ on this case, while AdaptiveGraph scores $0.0$, one
concrete instance of the diffusion penalty
(Section~\ref{sec:discussion-dilution}).

C3 (complete retrieval failure despite a helpful tag on every extraction
model) has no example because the category is empty in our failure subset
(Table~\ref{tab:failure-diagnostics}).

\end{document}